# Kinetic equations for Stark line shapes

J. Rosato, H. Capes, Y. Marandet, A. Mekkaoui, and R. Stamm

Laboratoire PIIM, UMR 7345 Université d'Aix-Marseille / CNRS, Centre de St-Jérôme,

Case 232, F-13397 Marseille Cedex 20, France

Phone +33-491-288624, Fax +33-491-670222

joel.rosato@univ-amu.fr

The BBGKY formalism is revisited in the framework of plasma spectroscopy. We address the issue of Stark line shape modeling by using kinetic transport equations. In the most simplified treatment of these equations, triple correlations between an emitter and the perturbing charged particles are neglected and a collisional description of Stark effect is obtained. Here we relax this assumption and retain triple correlations using a generalization of the Kirkwood truncature hypothesis to quantum operator. An application to hydrogen lines is done in the context of plasma diagnostic, and indicates that the neglect of triple correlations can lead to a significant overestimate of the line width.

---

1 – Introduction

The BBGKY formalism is usually involved in transport models for gases or plasmas (e.g. Balescu, 1997). In this paper we consider an application to the description of Stark line shapes for plasma spectroscopy purposes. This issue is of interest for diagnostics, e.g. in astrophysics, in fusion experiments or in arc discharges. The light naturally emitted from the atomic species (including multi-charged ions) contains information on the plasma parameters (N, $T_e$, etc.) and, hence, provides a valuable non-intrusive diagnostic tool. One of the most challenging theoretical issues in plasma spectroscopy concerns the development of Stark broadening models (Griem, 1974). Stark line shapes result from the interaction between the emitters and the microscopic electric field created from the charged particles. The models usually involve the solving of a statistical time-dependent quantum problem, given that the line shapes are proportional to the Fourier transform of the emitters' dipole autocorrelation function (see Griem, 1974). This quantity involves the statistical average of the evolution operator U(t) used in quantum mechanics, performed over the perturbers' states. Assuming classical kinetic plasma, these perturbers are characterized by the classical $\mathcal{N}$-particle phase space distribution $f_{\mathcal{N}}(1,\ldots,\mathcal{N};t)$. One of the most widely used models for Stark broadening at low density is the so-called impact approximation: it is assumed that the atom interacts briefly with one perturber at once, like collisions, and the resulting line shape is a Lorentzian function, whose width is determined from matrix elements of an operator K characterizing the collision frequency. In this work, we develop a generalization of the collision operator model suitable for multiple correlations. This is done in the framework of the so-called unified theory for Stark broadening (Voslamber, 1969; Smith et al. 1969). The general formalism is presented in Sec. 2, and the generalization of the binary model is done in Sec. 3. An application to hydrogen lines is done in the context of plasma diagnostic. The model is found in a good agreement with ab initio simulations. Furthermore, we show that the neglect of triple correlations can lead to a significant overestimate of the line width.

2 – Formalism

The formalism presented hereafter is usually referred to as "unified theory" in the literature on plasma spectroscopy. A detailed description in terms of the BBGKY hierarchy can be found in (Capes and Voslamber, 1972). In this approach, the system of interest is provided by an emitter immersed in a set of charged particles, and these particles are characterized by a classical $\mathcal{N}$-particle phase space distribution ("semi-classical approach"). The emitter is described quantum-mechanically and is usually referred to as an atom, even though it can be non-neutral. The spectral profile of a Stark broadened line is proportional to the Fourier transform of the emitter's dipole autocorrelation function (Griem, 1974). The latter quantity is calculated from matrix elements of the evolution operator, U(t), averaged over the plasma's microscopic electric field. This operator obeys the time-dependent Schrödinger equation (we set ℏ = 1)

$$i\frac{dU}{dt}(t) = (H_0 + V_{tot}(t))U(t), \qquad (1)$$

where $H_0$ is the Hamiltonian accounting for the atomic energy level structure and $V_{tot}(t) = -\vec{d} \cdot \vec{E}_{tot}(t)$ is the time-dependent Stark effect term resulting from the action of the total microscopic electric field $\vec{E}_{tot}(t)$, i.e. corresponding to the contribution of all the perturbers. The evolution operator is implicitly dependent on the plasma parameters through the phase space coordinates of the perturbers at the initial time. For convenience we will write U(t) ≡ U(1,…,𝒩;t) where 1,…,𝒩 is a shortcut notation for $\vec{r}_1, \vec{v}_1,…,\vec{r}_𝒩, \vec{v}_𝒩$. For the sake of simplicity we assume a one-component plasma of Debye quasiparticles. Following (Capes and Voslamber, 1972), we introduce a hierarchy of operators $\Phi_p(1,…,p;t)$, such that $\Phi_𝒩(1,…,𝒩;t) = f_𝒩(1,…,𝒩;t)U(t)$ where $f_𝒩$ is the 𝒩-particle phase space distribution and $\Phi_p(1,…,p;t) = \int d(p+1)…d𝒩 \Phi_𝒩(1,…,𝒩;t)$ for $0 \leq p \leq 𝒩$. These operators obey the following set of equations

$$\left\{\frac{\partial}{\partial t} + i\left[H_0 + \sum_{j=1}^{p} V(j)\right] + \sum_{j=1}^{p} \vec{v}_j \cdot \frac{\partial}{\partial \vec{r}_j}\right\} \Phi_p = -i(𝒩-p)\int d(p+1) V(p+1) \Phi_{p+1}, \quad (2)$$

with the initial conditions $\Phi_p(1,…,p;t=0) = f_p(1,…,p;t=0)$ (here $f_p$ refers to the p-particle reduced phase space distribution) and the convention $\Phi_{N+1} \equiv 0$. In Eq. (2), $V(j) = V(\vec{r}_j)$ denotes the Stark term resulting from the electric field due to the j-th perturber. It is assumed that the perturbers are independent and follow straight-line trajectories. Next, we focus on a uniform and stationary medium, so that $f_p(1,…,p;t) \equiv f_1(1)…f_1(p) \equiv f_1(\vec{v}_1)…f_1(\vec{v}_p)$.

The line shape is obtained from the operator $\Phi_0(t) = \int d1…d𝒩 \Phi_𝒩(1,…,𝒩;t)$, which is identical to the statistical average of the evolution operator. As in kinetic theory, it is practical to introduce a cluster representation of the $\Phi_p$'s (e.g. Balescu, 1997). These quantities are factorized as $\Phi_p(1,…,p;t) = f_p(1,…,p;t)\Phi_0(t)$ in the case where the atom and the perturbers do not interact with each other. If interactions are present, we define a set of generalized correlation functions $\Gamma_p(1,…,p;t)$ following this scheme (t is not written explicitly):

$$\begin{cases} \Phi_1(1) = f_1(1)\Phi_0 + \Gamma_1(1) \\ \Phi_2(1,2) = f_2(1,2)\Phi_0 + f_1(1)\Gamma_1(2) + f_1(2)\Gamma_1(1) + \Gamma_2(1,2) \\ \Phi_3(1,2,3) = f_3(1,2,3)\Phi_0 + f_2(1,2)\Gamma_1(3) + f_2(1,3)\Gamma_1(2) + f_2(2,3)\Gamma_1(1) \\ \qquad\qquad + f_1(1)\Gamma_2(2,3) + f_1(2)\Gamma_2(1,3) + f_1(3)\Gamma_2(1,2) + \Gamma_3(1,2,3) \\ \cdots \end{cases} \quad (3)$$

The $\Gamma_p$'s obey the following hierarchy of equations

$$\left\{\frac{\partial}{\partial t} + i\left[H_0 + \sum_{j=1}^{p}V(j)\right] + \sum_{j=1}^{p}\vec{v}_j \cdot \frac{\partial}{\partial \vec{r}_j}\right\}\Gamma_p(1,\ldots,p;t) =$$
$$-i\sum_{j=1}^{p}f_1(j)V(j)\Gamma_{p-1}(1,\ldots,j-1,j+1,\ldots,p;t) - i\mathcal{N}\int d(p+1)V(p+1)\Gamma_{p+1}(1,\ldots,p+1;t)$$
,(4)

with the initial conditions $\Gamma_p(1,\ldots,p;t=0) = 0$ for $p \geq 1$. In Eq. (4), by convention $\Gamma_0 \equiv \Phi_0$ and $\Gamma_{-1} \equiv 0$.

Equation (4) serves as a basis for a line shape calculation within the unified theory. Essentially, this theory leads to describe a spectral line as a sum of generalized Lorentzian functions, whose widths are frequency-dependent and given by a diagonal matrix element of an operator K characterizing the Stark perturbation. Mathematically, this corresponds to writing the Laplace transform of $\Phi_0$ as $\tilde{\Phi}_0(s) \equiv [s + iH_0 + K(s)]^{-1}$ where the Laplace variable s is identified as $-i\omega$, with $\omega$ being the photon frequency. The K-operator – usually referred to as "collision operator" – can be defined formally from Eq. (4) by setting p = 0, performing a Laplace transform, and multiplying the right-hand side by $1 = \tilde{\Phi}_0^{-1}(s)\tilde{\Phi}_0(s)$. Explicitly, one has

$$K(s) = i\mathcal{N}\int d1 V(1)\tilde{\Gamma}_1(1;s)\tilde{\Phi}_0^{-1}(s). \quad (5)$$

From this relation, it appears that determining the collision operator amounts to establish an expression for $\tilde{\Gamma}_1\tilde{\Phi}_0^{-1}$ which is independent both of $\Gamma_1$ and $\Phi_0$. In the original

BBGKY-approach to the unified theory (Voslamber, 1969), such an expression was obtained by neglecting multiple correlations. $\Gamma_1$ is obtained from solving Eq. (4) for p = 1 with $\Gamma_2 \equiv 0$, formally

$$\Gamma_1(1,t) = -if_1(1) \int_0^t d\tau Q(\vec{r}_1 - \vec{v}_1\tau, \vec{v}_1, \tau) V(\vec{r}_1 - \vec{v}_1\tau) \Phi_0(t-\tau), \tag{6}$$

where Q(1,t) is the propagator of the atom under the influence of one perturber. It obeys the time-dependent Schrödinger equation

$$\left\{ \frac{\partial}{\partial t} + i[H_0 + V(\vec{r}_1 + \vec{v}_1 t)] \right\} Q(1,t) = 0, \tag{7}$$

with the initial condition Q(1,t=0) = 1, and it is proportional to a time-ordered exponential (Dyson series)

$$Q(1,t) = e^{-iH_0 t} T \exp\left[ -i \int_0^t d\tau V(\vec{r}_1 + \vec{v}_1 \tau) \right], \tag{8}$$

T being the time-ordering operator. From Eq. (6), $\Gamma_1$ appears as a time convolution involving $\Phi_0$, so that its Laplace transform is proportional to $\tilde{\Phi}_0(s)$. Hence, the collision operator Eq. (5) is completely determined in terms of the propagator Q and the interaction term V. Algebraic manipulations lead to the following expression:

$$K(s) = \mathcal{N} \int_0^\infty dt e^{-st} \int d1 V(\vec{r}_1 + \vec{v}_1 t) Q(1,t) V(\vec{r}_1) f_1(1). \tag{9}$$

This is the main result of the unified theory. The broadening of a spectral line due to binary interactions is completely determined by the collision operator given in Eq. (9).

3 – Multiple collisions

We go beyond the binary approximation by assuming $\Gamma_2$ finite. A detailed treatment of the model developed hereafter can be found in (Capes, 1980). In order to get a simple expression for $\Gamma_2$, we write down Eq. (4) for p = 2 and we treat it by a perturbative approach assuming V small. At the second order, we obtain, for $\Gamma_2$,

$$\left\{\frac{\partial}{\partial t}+iH_0+\sum_{j=1}^{2}\vec{v}_j\cdot\frac{\partial}{\partial \vec{r}_j}\right\}\Gamma_2(1,2;t)=-if_1(1)V(1)\Gamma_1(2;t)-if_1(2)V(2)\Gamma_1(1;t). \quad (10)$$

This equation can be solved using a Green function method. A practical expression is obtained for the Fourier and Laplace transform of $\Gamma_2$

$$\tilde{\hat{\Gamma}}_2(\vec{k}_1,\vec{v}_1,\vec{k}_2,\vec{v}_2;s)=\tilde{\hat{\Gamma}}_1(\vec{k}_1,\vec{v}_1,s+i\vec{k}_2\cdot\vec{v}_2)\tilde{\Phi}_0^{-1}(s+i\vec{k}_2\cdot\vec{v}_2)\tilde{\hat{\Gamma}}_1(\vec{k}_2,\vec{v}_2;s)+1\leftrightarrow 2, \quad (11)$$

where the notation $\hat{F}(\vec{k})=\int d^3r e^{-i\vec{k}\cdot\vec{r}}F(\vec{r})$ has been used for any function of space $F(\vec{r})$ and where $1\leftrightarrow 2$ denotes permutation between the variables. Equation (11) presents similarities with the truncature hypothesis proposed by Kirkwood in kinetic theory (Kirkwood, 1935), where here $\Gamma_1$ and $\Phi_0$ play the role of the two-particle correlation function $g_2$ and the one-particle distribution function $f_1$, respectively. A closed evolution equation for $\Gamma_1$ is obtained by inserting the inverse Fourier and Laplace transform of Eq. (11) into Eq. (4) with p = 1. Following (Capes, 1980), we obtain an expression similar to that obtained within the binary approximation [Eq. (6)] in terms of an effective propagator $Q_{eff}(1,t)$ accounting for multiple collisions. Namely,

$$\Gamma_1(1,t)=-if_1(1)\int_0^t d\tau Q_{eff}(\vec{r}_1-\vec{v}_1\tau,\vec{v}_1,\tau)V(\vec{r}_1-\vec{v}_1\tau)\Phi_0(t-\tau). \quad (12)$$

$Q_{eff}(1,t)$ satisfies the following equation

$$\left\{\frac{\partial}{\partial t}+i[H_0+V(\vec{r}_1+\vec{v}_1t)]\right\}Q_{eff}(1,t)+\int_0^t d\tau M(\tau)Q_{eff}(1,t-\tau)=0, \quad (13)$$

with the initial condition $Q_{eff}(1,t=0) = 1$. It can be interpreted as describing the evolution of the atom under the influence of one collision represented by the interaction term V, given a set of collisions occurring in its past history. These collisions are characterized by the kernel M(t). This term is identical to the inverse Laplace transform of the collision operator. In the case where multiple collisions are neglected, $M \equiv 0$ and Eq. (13) becomes identical to the Schrödinger equation used in the binary approximation, Eq. (7). Note, Eq. (12) is an approximation of the formal solution of Eq. (4) with p = 1, suitable for numerical calculations. The exact formal solution involves a double time integral (Capes, 1980) and will be investigated in details in a future work.

The collision operator is obtained from its formal definition Eq. (5). An explicit calculation leads to

$$K(s) = \mathcal{N} \int_0^\infty dt e^{-st} \int d1 V(\vec{r}_1 + \vec{v}_1 t) Q_{eff}(1,t) V(1) f_1(1). \tag{14}$$

This equation has a structure similar to the expression of the collision operator within the binary model, Eq. (9). The presence of multiple collisions is retained through the term $Q_{eff}$. In contrast with the binary case, this is not a closed expression because $Q_{eff}$ depends on the collision operator. It is determined from the evolution equation (13), which involves the inverse Laplace transform of K. In practice, a calculation should be done by iterations. The additional term in Eq. (13) denotes a resonance damping. This is illustrated by rewriting Eq. (13) in the Fourier and Laplace domain:

$$\begin{aligned}\tilde{\hat{Q}}_{eff}(\vec{k}_1,\vec{v}_1,s) &= (2\pi)^3 \delta(\vec{k}_1)[s+iH_0+K(s)]^{-1} \\ &- \frac{i}{(2\pi)^3}[s+iH_0+K(s)]^{-1} \int d^3 k_2 \hat{V}(\vec{k}_2)\tilde{\hat{Q}}_{eff}(\vec{k}_1-\vec{k}_2,\vec{v}_1,s-i\vec{k}_2\cdot\vec{v}_1)\end{aligned} \tag{15}$$

As can be seen, the collision operator appears in the denominator as a non-Hermitian contribution to the Hamiltonian, which attenuates the resonance. This attenuation implies a reduction of the Stark effect present during one binary interaction by the presence of other perturbers. In practice, these perturbers reduce the range of the electric field involved in a binary interaction in a fashion similar to the Debye screening, with the characteristic length v/γ where γ is a typical matrix element of the collision operator. This length can be interpreted as the mean free path of an atom between two collisions. Therefore, multiple collisions are expected to be relevant when v/γ becomes of the same order as the Debye length or smaller. Equation (15) presents similarities with the result of the so-called resonance broadening theory used for plasma turbulence (Dupree, 1966; Weinstock, 1969), where the quasi-linearized Vlasov equation plays a role similar to Eq. (13). In this theory, the coupling between the one-particle distribution function and the plasma's electric field is described through a diffusion coefficient in the velocity space and the latter obeys a nonlinear equation as does our collision operator.

A simplification, practical for numerical calculations, is provided by assuming K(s) ≈ K(-i$\omega_0$) ≡ $K_0$ in Eq. (15), using that the collision operator is governed by the values of $\tilde{\tilde{Q}}_{eff}$ near the resonance. This amounts to setting M(t) ≡ $K_0\delta(t)$ in Eq. (13), and it leads to a simple expression for $Q_{eff}$, with a structure similar to that in the binary case Eq. (8):

$$Q_{eff}(1,t) = e^{-i(H_0 - iK_0)t} T \exp\left[-i\int_0^t d\tau V(\vec{r}_1 + \vec{v}_1\tau)\right]. \tag{16}$$

We have applied the collision operator formula Eq. (14) to calculations of hydrogen line shapes in an ideal case. The collision operator has been calculated by iterations. Figure 1 presents a plot of the Lyman α line (n = 2 → 1) broadened due to ions at N = 2.5×10$^{15}$

cm$^{-3}$, T = 1.1 eV, obtained using the unified theory (binary approximation) compared to that obtained within the $Q_{eff}$-model. The ratio $v/<210|K_0|210>\lambda_D$ is of the order of unity, which means that non-binary interactions are expected to be relevant. A benchmark result from an ab initio simulation code (Rosato et al. 2009) is also shown in the figure. As can be seen, the binary model overestimates the width by 40% whereas the $Q_{eff}$-model gives a much better result, with an overestimate of 10%. This residual overestimate can be due to an oversimplification of the collision operator formula Eq. (14) used in the numerical procedure.

4 – Conclusion

The BBGKY formalism is suitable for plasma spectroscopy models. In this work, we have developed an extension of the so-called unified theory for Stark line shapes. Our model retains triple correlations between an emitter and the perturbing charged particles through an effective propagator in the collision operator, which accounts for the presence of multiple emitter-perturber interactions during a given collision. This treatment is similar to that used in the resonance broadening theory for plasma turbulence. The multiple interactions attenuate the Stark effect present during one binary interaction, so that a line shape is narrower than expected from a binary treatment. In a diagnostic context, this means that a reliable interpretation of spectroscopic observations should involve a careful examination of the role of non-binary interactions. Such an examination can be done using the theory presented in this work. In the plasma conditions considered above, our model yields results much closer to ab initio simulations than the binary approximation. A complement to this work should consist in addressing the limits of this model, e.g., by performing similar comparisons in other conditions.

Acknowledgements

This work is partially supported by the French National Research Agency (contract ANR-07-BLAN-0187-01) and by the collaboration (LRC DSM99-14) between the PIIM laboratory and the CEA Cadarache (Euratom Association) within the framework of the French Research Federation on Magnetic Fusion (FR-FCM).

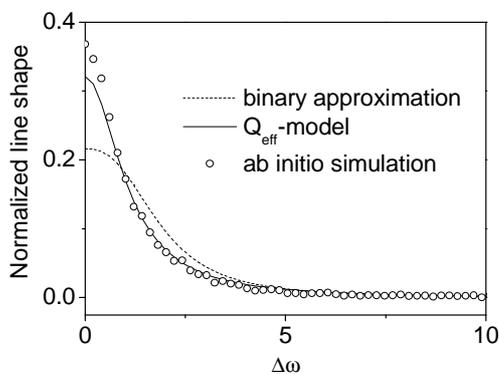

Figure 1 – Ion Stark broadening of the hydrogen Lyman α line obtained using a binary approximation (dashed line) compared to that obtained within the model developed in this work (solid line, referred to as "$Q_{eff}$-model"). Also shown in the figure is the result of an ab initio simulation code (circles), which serves as a benchmark. The $Q_{eff}$-model gives

a result closer to the simulation. In the x-axis, the frequency detuning Δω is in reduced units.